\documentclass[twocolumn,byrevtex,prl,showpacs]{revtex4}

\usepackage{graphicx}
\usepackage{dcolumn}
 
\begin{document}

\title{Ground-state phase diagram of a half-filled one-dimensional
extended Hubbard model}

\author{Eric Jeckelmann}
\affiliation{Fachbereich Physik, Philipps-Universit\"{a}t, 
D-35032 Marburg, Germany}

\date{\today}

\begin{abstract}
The density-matrix renormalization group is used to study the 
phase diagram of the one-dimensional half-filled Hubbard 
model with on-site ($U$) and nearest-neighbor ($V$) repulsion, and 
hopping $t$.  
A critical line $V_c(U) \approx U/2$ separates a Mott insulating phase 
from a charge-density-wave phase.
The formation of bound charge excitations for $V > 2t$
changes the phase transition from continuous to
first order at a tricritical point $U_t\approx 3.7t, V_t=2t$. 
A frustrating effective antiferromagnetic spin coupling induces a 
bond-order-wave phase on the critical line $V_c(U)$
for $U_t < U \alt 7t$.
\end{abstract}


\maketitle

The properties of quasi-one-dimensional materials have been extensively
studied in recent years \cite{kiess,bourbonnais,kishida}.
These materials exhibit rich phase diagrams and display unusual
optical properties due to the combination of reduced dimensionality
and strong electronic correlations. 
Consequently, much effort has been devoted to understanding the 
ground-state and optical properties of theoretical one-dimensional 
correlated electron systems such as the half-filled Hubbard 
model with on-site ($U$) and nearest-neighbor ($V$) repulsion and 
hopping term $t$.
Nevertheless, the ground state phase diagram of this model is still
controversial 
\cite{hirsch,cannon,peter,kampf,nakamura,tsuchiizu,sengupta}. 
It is known \cite{hirsch} that the system is a Mott insulator for $U 
\agt 2V$ and a charge density wave (CDW) insulator for $U \alt 2V$.  
The quantum phase transition is continuous at weak coupling
($U,V \ll t$) and first order at strong coupling ($U,V \gg t$).
Numerically \cite{hirsch,cannon,sengupta}, one finds that the order of 
the transition changes at a tricritical point ($U_t,V_t$) with 
$V_t/t \approx 1.5-2.5$, but this feature 
is not well understood.
Recently, it has been proposed \cite{nakamura,tsuchiizu,sengupta} 
that a  bond-order-wave (BOW) phase exists between the Mott and CDW 
phases up to the tricritical point.  
Concurrently, the optical properties of this model 
have been determined in the Mott insulating phase 
\cite{florian,shuai,controzzi,fabian,eric}.
In particular, it has been found that the lowest optical excitations
consist of a pair of independent charge excitations
for $V \leq 2t$, while they are bound states for $V > 2t$.
Surprisingly, the remarkable proximity of the tricritical point to the 
boundary between bound and free charge excitations has not been 
noticed until now. 

Here I investigate the ground-state phase diagram of the half-filled 
one-dimensional extended Hubbard model in the repulsive regime 
using the density-matrix renormalization group (DMRG) \cite{steve}.
I show that the nature of the low-lying charge excitations determines 
the order of the transition and the position of the tricritical point.  
A BOW phase is found only at intermediate coupling
on the critical line $V_c(U)$ between Mott and CDW phases.

The model is defined by the Hamiltonian
\begin{eqnarray}
\hat{H} 
&=& -t \sum_{l;\sigma} \left( \hat{c}_{l,\sigma}^+\hat{c}_{l+1,\sigma} 
+ \hat{c}_{l+1,\sigma}^+\hat{c}_{l,\sigma} \right) \nonumber \\
&&+ U \sum_{l} \left(\hat{n}_{l,\uparrow}-\frac{1}{2}\right)
\left(\hat{n}_{l,\downarrow}-\frac{1}{2}\right)  \nonumber \\
&& + V \sum_{l}(\hat{n}_l-1)(\hat{n}_{l+1}-1)  \; .
\label{hamiltonian}
\end{eqnarray}
Here $\hat{c}^+_{l,\sigma}$, $\hat{c}_{l,\sigma}$ are creation and 
annihilation operators for electrons with spin 
$\sigma = \uparrow,\downarrow$ at site $l=1,\dots,N, \hat{n}_{l,\sigma}=
\hat{c}^+_{l,\sigma}\hat{c}_{l,\sigma}$, 
and $\hat{n}_l=\hat{n}_{l,\uparrow}+\hat{n}_{l,\downarrow}$.
I exclusively consider systems with an even number $N$ of sites.
At half filling the number of electrons equals $N$.
The interaction is repulsive $U \geq V \geq 0$.
The Hamiltonian~(\ref{hamiltonian}) has a particle-hole symmetry and a 
spatial reflection symmetry. 
Thus, each eigenstate has a well-defined parity under charge 
conjugation ($P_{c} = \pm 1$) and belongs to one of the two irreducible
representations, $A_{g}$ or $B_{u}$, of the reflection symmetry group.  
The ground state belongs to the symmetry subspace 
$A_{g}^{+} \equiv (A_{g},P_{c})$ and optically excited states belong to
the symmetry subspace $B_{u}^{-} \equiv (B_{u},-P_{c})$ because   
the current operator is antisymmetric under charge conjugation and 
spatial reflection \cite{fabian}. 

DMRG~\cite{steve}
is known to be a very accurate numerical method for one-dimensional
quantum systems with short-range interactions such as
the Hamiltonian~(\ref{hamiltonian}).
Here I use the finite-system DMRG algorithm to calculate
eigenenergies and static properties at low-energy.  
Spin gaps $E_{s} = E_0(N,1) -  E_0(N,0)$ and single-particle charge gaps
$E_{c} = 2  [ E_0(N+1,1/2) - E_0(N,0) ]$ are derived from
the ground state energies $E_0(N_e,S_z)$ for a given number of electron 
$N_e$ and a given spin $S_z$. 
I also use a symmetrized DMRG~\cite{ramasesha,eric} method to 
calculate the energy and static properties of the lowest eigenstates 
in the $B_u^-$ symmetry subspace.

All DMRG methods have a truncation error which is 
reduced by increasing the number~$m$ of density-matrix 
eigenstates kept in the renormalization procedure 
\cite{steve}. 
To achieve a greater accuracy and to obtain error estimates 
I extrapolate DMRG results to the limit of 
vanishing discarded weight $P_m$.
For eigenenergies DMRG errors vanish linearly with $P_m$,
while for other quantities DMRG errors usually 
scale as $(P_m)^{\alpha}$ with $0 < \alpha \leq 1$.
The largest number of density-matrix eigenstates used
in this work is $m=1200$ and truncation errors are negligible
for all results presented here. 
DMRG calculations have always been carried out for several sizes  
$N$ in order to check finite-size effects 
and, if necessary, results have been extrapolated to the thermodynamic 
limit $N \rightarrow \infty$. 
The largest system size used in this work is $N=1024$. 

\begin{table}
\caption{ \label{table1}
Critical nearest-neighbor repulsion
$V_{c}(U)$, single-particle charge gap $E_c$ at ($U,V_c$),
and transition type for several values of $U$.
Numbers in parenthesis are estimated errors.
}
\begin{ruledtabular}
\begin{tabular}{rddc}
$U/t$ & \multicolumn{1}{r}{$V_c(U)/t \; \;$} & 
\multicolumn{1}{r}{$E_c/t$} & transition order\\
\colrule    
2& 1.125~(25)& 0 & continuous\\
3& 1.640~(10)& 0 & continuous\\
4& 2.150~(10)& 0.05 & first order\\
5& 2.665~(5)& 0.11 & first order\\
6& 3.155~(5)& 0.15 & first order\\
8& 4.141~(4)& 0.45 & first order\\
12& 6.115~(5)& 2.0 & first order\\
40& 20.041~(4)& 29.3 & first order\\
\end{tabular}
\end{ruledtabular}
\end{table}

For $V \alt U/2$ the model (\ref{hamiltonian}) describes
a Mott insulator.
[This phase is often called a spin-density-wave (SDW) phase because 
antiferromagnetic spin correlations decay algebraically in the ground 
state.] 
There is no broken symmetry and the ground state is nondegenerate. 
The charge gap $E_c$ is finite but the spin gap $E_s$ vanishes 
in the thermodynamic limit. 
In this Mott insulating phase the lowest eigenstate in the $B_u^-$ 
symmetry sector always contributes to the optical spectrum.
Therefore, the difference between the $B_u^-$ and $A_g^+$ 
ground-state energies corresponds to the optical gap $E_{\text{opt}}$.
For $V \agt U/2$ the system is in a long-range ordered CDW phase
with a doubly degenerate ground state in the thermodynamic limit.
In this phase both charge and spin gaps are finite.
The CDW order parameter $0 < |m_e| \leq 1$ gives the amplitude
of the ground-state density modulation 
$\langle \hat{n}_l \rangle = 1 + (-1)^l m_e$.          
Such a CDW breaks the charge-conjugation symmetry and the spatial
reflection symmetry.
Thus one of the two degenerate ground states belongs to the 
$A_g^+$ symmetry subspace and the other one to the $B_u^-$ subspace.
Obviously, this consideration shows that the low-energy $B_u^-$ 
excitations in the Mott phase play an important role in the 
Mott-CDW transition.

The transition from the CDW phase to the Mott phase can be investigated
with DMRG as done previously for the CDW-metal transition in the 
Holstein model~\cite{eric3}.
More precisely, I have investigated the lowest 
eigenstates to determine the ground-state degeneracy and symmetry, and
I have calculated the electronic staggered susceptibility to check if 
the ground state has long-range CDW order.
Some results for the critical line $V_{c}(U)$ determined with this 
approach are given in Table~\ref{table1}.
These results agree quantitatively with recent quantum Monte Carlo 
(QMC) simulations \cite{sengupta} and the Gaussian transition line in 
Ref.~\onlinecite{nakamura}, 
although the interpretation of this phase boundary is completely 
different (see below). 
These values are also in good agreement with strong-coupling 
perturbation theory \cite{peter} down to $U=6t$ 
and the results of early numerical 
simulations \cite{cannon,hirsch}.
This confirms the accuracy of the present approach 
and I will not elaborate further on the location of this boundary. 

\begin{figure}
\includegraphics[width=6cm]{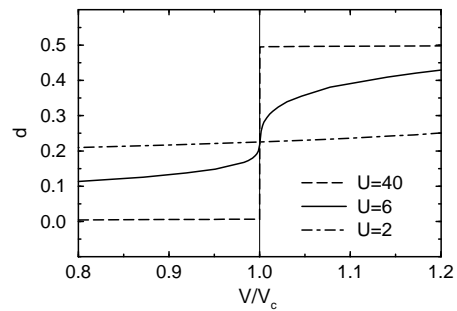}
\caption{ \label{fig1}
Ground-state double occupancy $d$ versus $V$ close to the critical 
coupling $V_{c}(U)$ for several $U$ (in units of $t$).
}
\end{figure}

To determine the order of the transition on the critical line
$V_{c}(U)$, one can examine the derivatives of the ground-state energy
per site with respect to the interaction parameters $U$ and $V$.
Using the Hellmann-Feynman theorem one shows that the derivatives
are given (up to known constants) by ground-state expectation values of
the double occupancy and nearest-neighbor 
density-density operators, respectively. 
These expectation values can be calculated accurately with DMRG.
Figure~\ref{fig1} shows the average ground-state double occupancy
\begin{equation}
d =  \frac{1}{N}  \sum_{l} \langle \hat{n}_{l \uparrow} 
\hat{n}_{l \downarrow} \rangle
\label{doublondensity}
\end{equation}
versus $V$ close to the critical coupling $V_{c}(U)$.
For $U=40t$, $d$ jumps from almost zero to slightly less than $0.5$
at $V_c$ and thus the transition is clearly first order,
while it is continuous for $U=2t$. 
For $U=6t$ the derivative of $d$ with respect to $V$ evolves into
a $\delta$-function at $V = V_c$ as the system size $N$ increases.
Therefore, $d$ is discontinuous at $V_c(U)$ and the transition is also 
first order for $U=6t$.
With this method I have found that the transition at $V_c(U)$
is first order from 
the strong-coupling limit ($U,V \gg t$) down to at least 
$U=4t \ (V_c \approx 2.15t)$ and is 
continuous from the weak-coupling limit ($U,V \ll t$) up to at least 
$U=3.5t \ (V_c \approx 1.9t)$.

\begin{figure}
\includegraphics[width=6cm]{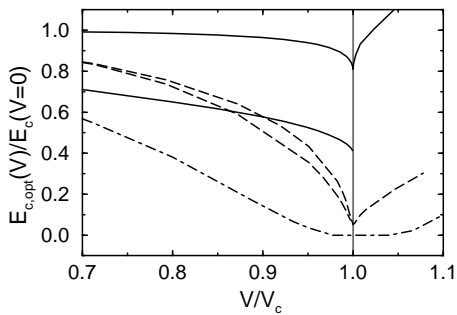}
\caption{\label{fig2}
Charge gap $E_c$ (upper line) and optical gap $E_{\text{opt}}$ 
(lower line) versus $V$ close to the critical coupling $V_c(U)$
for $U/t=40$ (solid), $6$ (dashed), and $2$ (dot-dashed).
For $U=2t$, $E_{\text{opt}} = E_c$.
$E_{\text{opt}}$ is not shown in the CDW phase ($V > V_c$).
}
\end{figure}

An investigation of the low-lying charge excitations in the 
Mott insulating phase allows us to understand the existence of a 
tricritical point and to determine its position.
Figure~\ref{fig2} shows the evolution of the charge gap $E_c$ and the
optical gap $E_{\text{opt}}$ for three different couplings $U$.
I have always found that the charge gap $E_c$ varies continuously and 
goes through a minimum at $V_c(U)$ as $V$ is increased for fixed $U$. 
For weak coupling ($U \leq 3t$) the gap vanishes smoothly on 
the critical line. For stronger coupling ($U \geq 4$)
the slope of $E_c$ with respect to $V$ becomes very large
and is discontinuous at $V_c$.
In this regime the charge gap remains finite at the transition
(see Table~\ref{table1}).

In a Mott insulator, elementary charge excitations can be understood
as spinless bosons in Hubbard bands.  
Excited states in the $B_u^-$ subspace always consist
of an even number of such elementary excitations with zero total
charge \cite{florian,controzzi,fabian}.
DMRG calculations combined with analytical methods \cite{fabian, eric}
reveal their properties.
For $V \leq 2t$ the low-energy $B_u^-$ excited states 
consist of two independent elementary charge excitations 
and the optical gap equals the charge gap in the thermodynamic limit.
For $V > 2t$, however, the lowest $B_u^-$ excitations
are bound states starting at an energy $E_{\text{opt}} < E_c$
(see Fig.~\ref{fig2}).
Just above $2t$ they are excitons made of two elementary charge 
excitations.  
Close to the critical line $V_{c}(U) \approx U/2$ 
they become finite-size ``droplets'' of the CDW phase,
(i.e., a bound state of several elementary charge excitations).
The excitation energy $E_{\text{opt}}$ of the lowest CDW droplet 
remains finite at the transition $V_c(U)$ at least for $U \geq 5$
(see also Fig.~\ref{fig2}).
CDW droplet sizes increase sharply as $V$ approaches the critical
coupling $V_c(U)$ for fixed $U$
and reach a finite value $\xi_c$ at the critical line.
This critical droplet size $\xi_c$ diverges for $U \rightarrow \infty$
but tends to 2 (corresponding to an exciton) if $V_c$ approaches $2t$. 

Combining all results, one discerns two distinct regimes
in the Mott phase as the critical line $V_c(U)$ is approached.
If $V_c \leq 2t$, the transition is continuous, the 
$B_u^-$ excitations are made of free charge excitations
and become gapless at $V_c$.
If $V_c > 2t$, the transition is first order,
the low-lying $B_u^-$ excitations are bound states and
remain gapped at $V_c$.
Therefore, I believe that the tricritical point is precisely
located at the 
intersection of the $V=2t$ boundary with the quantum critical line 
$V_c(U) \approx U/2$, i.e., at $U_t \approx 3.7t, V_t=2t$.

Adapting Hirsch's analysis \cite{hirsch}, one can now qualitatively 
explain why the transition from the Mott to the CDW phase
changes from first order to continuous. 
If $V_c(U)> 2t$, CDW droplets with sizes $\xi \geq \xi_c$ 
are energetically favored when $V$ becomes larger than $V_c(U)$, and 
the system tunnels to the CDW phase by nucleation.  
On the critical line $V_c(U)$ the creation energy $E_{\text{opt}}$
and critical size $\xi_c$ of the lowest
CDW droplet diminishes with $U$.
At $V_c=2t$, CDW droplets (with sizes $\xi \rightarrow 2$) become 
instable and split into independent gapless elementary charge 
excitations. 
For $V_c(U) \leq 2t$, it is then advantageous to create many
such excitations as $V$ is increased above $V_c(U)$
and the transition becomes continuous. 

In the field theory approach to the lattice model~(\ref{hamiltonian})
the limit between free and bound charge excitations corresponds to a 
parameter $K_{\rho}=1/2$ for a half-filled system 
\cite{controzzi,fabian}.
Field-theoretical predictions for the low-energy optical spectrum 
agree perfectly with dynamical DMRG calculations \cite{fabian,eric}. 
Thus the tricritical point probably corresponds to $K_{\rho}=1/2$.
In Ref.~\onlinecite{sengupta}, it has also been found that 
$K_{\rho}$ is close to $1/2$ in this region.

Recently, Nakamura~\cite{nakamura} has proposed that 
a narrow region with long-range BOW order exists between the 
CDW phase and the Mott phase at weak to intermediate coupling.
The BOW region vanishes at the tricritical point,
beyond which there is a 
direct first-order transition from the Mott phase to the CDW phase. 
In this scenario the critical line $V_c(U)$ calculated above
would correspond to the BOW-CDW phase boundary for $U < U_t$.
The existence of long-range BOW order in the critical region
has been confirmed by QMC simulations \cite{sengupta}. 

In a BOW phase the ground state is two-fold degenerate 
and both charge and spin gaps are finite. 
A long-range ordered BOW corresponds to a finite staggered bond 
order 
\begin{equation}
\frac{1}{2} \sum_{\sigma} \left \langle 
\hat{c}_{l,\sigma}^+\hat{c}_{l+1,\sigma} 
+ \hat{c}_{l+1,\sigma}^+\hat{c}_{l,\sigma} 
\right \rangle = p_0 + (-1)^l \delta 
\label{bondorder}
\end{equation}
with $\delta \neq 0$ in the thermodynamic limit.
To distinguish the BOW phase from the Mott phase I have calculated the
spin gap $E_s$ and the bond order parameter $\delta$ using DMRG.
After extrapolation to vanishing truncation errors $P_m$
and to the thermodynamic limit $N \rightarrow \infty$,
this approach gives spin gaps with an accuracy of $10^{-3}t$ or better
and allows one to detect a bond order alternation as small 
as $|\delta|=0.01$. 
(It appears numerically that $\delta$ 
either vanishes as $1/\sqrt{N}$ or tends to a finite value
with finite-size corrections scaling as $1/N$.) 
I have found a BOW ground state 
only in a very narrow region adjacent to the critical line
$V_{c}(U)$  for intermediate coupling $4t \leq U \leq 6t$.
For $U=4t$ and $V=2.14t$, I have obtained  $\delta = 0.08$ in 
quantitative agreement with QMC simulations \cite{sengupta}.
However, the BOW phase extends to significantly stronger coupling than 
reported in Ref~\onlinecite{sengupta}.  
For instance, there is a BOW ground state with $\delta\approx0.12$
for $U=6t$ and $V=3.145t$. 
I have not observed any BOW ground state at weak coupling ($U \leq 3t$),
at strong coupling ($U \geq 8t$), and for $V \leq U/2$. 
Actually, as BOW ground states are found only within $0.02t$ of the
critical line $V_c(U)$, I believe that the BOW phase exists only
\textit{on} this critical line for intermediate coupling $U$ starting
from the tricritical point $U_t \approx 3.7t$ up to a upper limit
which is smaller than $8t$.
Field theory \cite{kampf} also suggests that a BOW phase 
can occur only on the boundary between CDW and Mott phases in the 
model~(\ref{hamiltonian}).
Even in the Hartree-Fock approximation, one finds a BOW
phase only on the critical line $U=2V$ between a SDW 
phase and a CDW phase \cite{dionys}.

The occurrence of the BOW phase on the critical line for $U \alt 8t$
can be understood as the result of increasing frustration in the
spin degrees of freedom.
On this critical line, 
low-energy charge excitations are dispersionless and thus ineffective
in a degenerate CDW-Mott ground state.
A strong-coupling expansion up to 4th order in $t/U$ shows that
the spin properties are determined by an effective
Heisenberg Hamiltonian with nearest- ($J_1$) and 
next-nearest-neighbor ($J_2$) antiferromagnetic exchange couplings
\cite{peter}.
This strong-coupling perturbation analysis gives accurate results for 
the critical line $V_c(U)$ (and thus is expected to be valid)
down to $U=6t$.  
The ratio $J_2/J_1$ is strongly enhanced by the nearest-neighbor 
repulsion $V$.
It is known \cite{robert,steve2} that the frustration due to the $J_2$ 
coupling causes a quantum 
phase transition from a ``spin-liquid'' ground state for $4J_2 \alt J_1$
to a dimerized spin ground state for $4J_2 \agt J_1$ in the Heisenberg 
model.  
On the critical line $V_c(U)$ of the extended Hubbard model, 
this corresponds to the appearance of a BOW phase (driven by the spin 
dimerization) for $U \alt 7t$, in agreement with my DMRG results,
if one uses the values of $J_1$ and $J_2$
given by the 4th-order perturbation expansion \cite{peter}.
Away from the critical line [$V < V_c(U)$], one observes only
algebraically decreasing BOW correlations because charge fluctuations,
which are no longer dispersionless, destroy the local spin moments
and thus prevent any long-range spin order.

\begin{figure}
\includegraphics[width=6cm]{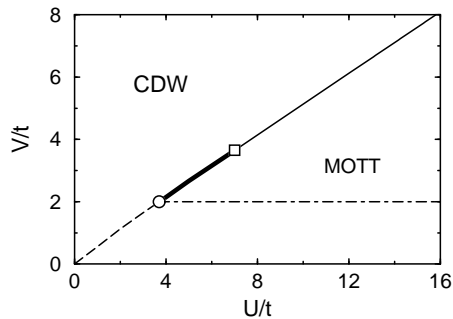}
\caption{ \label{fig3}
Schematic ground-state phase diagram: the transition is 
continuous at weak coupling (dashed line)
and first order at strong coupling (both solid lines).
A circle marks the tricritical point. 
The dot-dashed line is the boundary between free and bound charge 
excitations in the Mott phase.
The thick solid line indicates the BOW phase and 
a square marks the transition from a ``spin-liquid''
to a dimerized spin ground state on the critical line.}
\end{figure}

Figure~\ref{fig3} shows the phase diagram found with DMRG.
This phase diagram disproves recent speculations about an extended
BOW phase from weak coupling up to the tricritical point 
\cite{nakamura,sengupta,tsuchiizu} but
is compatible with the results of most previous investigations
of this problem (see Refs.~\onlinecite{hirsch} to~\onlinecite{kampf} 
and references therein).  

In summary, I have studied the ground-state phase diagram
of the one-dimensional half-filled extended Hubbard model
for $U,V > 0$ using DMRG.
I have shown that the phase diagram is determined by 3 mechanisms: 
First, the competition between the on-site ($U$)
and nearest-neighbor ($V$) repulsion is responsible for a quantum 
critical transition from a Mott to a CDW ground state at $V\approx U/2$.
Second, the competition between the hopping term $t$ and 
the nearest-neighbor repulsion $V$ is responsible for the 
transformation of free charge excitations into bound states 
(CDW droplets) at $V=2t$ in the Mott phase 
and thus changes the Mott-CDW transition from continuous to first order.
Third, a frustrating effective antiferromagnetic spin coupling
leads to a dimerized spin ground state (and thus to a BOW phase)
on the Mott-CDW critical line at intermediate coupling.



\begin{thebibliography}{99}

\bibitem{kiess} \textit{Conjugated Conducting Polymers}, edited
by H.~Kiess (Springer, Berlin, 1992).

\bibitem{bourbonnais} C. Bourbonnais and D. J\'erome, in
\textit{Advances in Synthetic Metals, Twenty Years of Progress
in Science and Technology}, edited by P. Bernier, S. Lefrant, and
G. Bidan (Elsevier, New York, 1999), pp. 206-301.

\bibitem{kishida} H. Kishida \textit{et al.}, Nature (London) 
\textbf{405}, 929 (2000).

\bibitem{hirsch} J.E.~Hirsch, \prl \textbf{53}, 2327 (1984).

\bibitem{cannon} J.W. Cannon, R.T. Scalettar, and E. Fradkin,
\prb \textbf{44}, 5995 (1991).

\bibitem{peter} P.G.J. van Dongen, \prb \textbf{49}, 7904 (1994).

\bibitem{kampf} G.I. Japaridze and A.P. Kampf, \prb \textbf{59}, 12822 
(1999). 

\bibitem{nakamura} M. Nakamura, \prb \textbf{61}, 16377 (2000).

\bibitem{tsuchiizu} M. Tsuchiizu and A. Furusaki, \prl \textbf{88},
056402 (2002).

\bibitem{sengupta}  P. Sengupta, A.W.~Sandvik, and D.K.~Campbell,
\prb \textbf{65}, 155113 (2002). 

\bibitem{florian} F.~Gebhard, K.~Bott, M.~Scheidler, P.~Thomas, and 
S.W.~Koch, Philos.~Mag.~B~\textbf{75}, 47 (1997).

\bibitem{shuai} Z.~Shuai, J.L.~Br\'edas, S.K.~Pati,  
and S.~Ramasesha, \prb~\textbf{58}, 15329 (1998).

\bibitem{controzzi} D. Controzzi, F.H.L. Essler, and A.M. Tsvelik, 
\prl \textbf{86}, 680 (2001). 

\bibitem{fabian} F.H.L.~Essler, F.~Gebhard, and E.~Jeckelmann, \prb
\textbf{64}, 125119 (2001). 

\bibitem{eric} E.~Jeckelmann, e-print cond-mat/0208480 (unpublished).

\bibitem{steve} S.R.~White, \prl~\textbf{69}, 2863 (1992); 
\prb~\textbf{48}, 10345 (1993).

\bibitem{ramasesha} 
S.~Ramasesha, S.K.~Pati, H.R.~Krishnamurthy,
Z.~Shuai, and J.L.~Br\'edas, Phys.~Rev.~B.~\textbf{54}, 7598 (1996).


\bibitem{eric3} E. Jeckelmann, C. Zhang, and S.R. White, 
\prb \textbf{60}, 7950 (1999).

\bibitem{dionys} D. Baeriswyl in \textit{Theoretical Aspects of Band
Structures and Electronic Properties of Pseudo-One-Dimensional
Solids}, edited by R.H. Kamimura (Reidel, Dordrecht, Holland, 1985),
pp. 1-48.

\bibitem{robert} R. Bursill \textit{et al.}, 
J. Phys. Condens. Matter \textbf{7}, 8605 (1995).

\bibitem{steve2} S.R. White and I. Affleck, \prb \textbf{54}, 9862 
(1996).

\end{thebibliography}
\end{document}